\documentclass[aps,twocolumn,superscriptaddress,noshowkeys,amssymb,amsmath,amsfonts,longbibliography]{revtex4-1}

\usepackage{graphicx}
\usepackage{theorem}
\usepackage{dcolumn}
\usepackage{amssymb}
\usepackage{amsmath}
\usepackage{amsfonts}
\usepackage{bm}
\usepackage{tikz}
\usepackage{epstopdf}
\usepackage{soul}

\newcommand{\SSS}{\scriptscriptstyle}
\newcommand{\DS}{\displaystyle}

\newcommand{\Ee}{{\rm e}}
\newcommand{\Dd}{{\rm d}}

\newcommand{\Ii}{{\rm i}}

\newcommand{\Ex}{\mathbf{e}_x}
\newcommand{\Ey}{\mathbf{e}_y}

\newcommand{\Ez}{\mathbf{e}_z}

\newcommand{\EpsM}{\epsilon_{{\SSS\text{M}}}}
\newcommand{\EpsA}{\epsilon_{{\SSS\text{A}}}}
\newcommand{\EpsB}{\epsilon_{{\SSS\text{B}}}}

\newcommand{\DdA}{d_{\SSS\text{A}}}
\newcommand{\DdB}{d_{\SSS\text{B}}}

\newcommand{\calJ}{j}

\newcommand{\vP}{v_{\text{p}}}
\newcommand{\JD}{j_{\SSS\text{D}}}
\newcommand{\JR}{j_{\SSS\text{R}}}
\newcommand{\JL}{j_{\SSS\text{L}}}
\newcommand{\JLR}{j_{\SSS\text{L,R}}}

\newcommand{\OmegaC}{\omega_{\text{c}}}
\newcommand{\OmegaP}{\omega_{\text{p}}}

\newcommand{\Nabla}{\nabla_{\bm{r}}}

\newcommand{\EF}{E_{\SSS\text{F}}}

\begin{document}

\title{Topological magnetoplasmon}

\author{Dafei Jin}
\affiliation{Department of Mechanical Engineering, Massachusetts Institute of Technology, Cambridge, Massachusetts 02139, USA}
\author{Ling Lu}\email{linglu@iphy.ac.cn}
\affiliation{Institute of Physics, Chinese Academy of Sciences/Beijing National Laboratory for Condensed Matter Physics, Beijing 100190, China}
\affiliation{Department of Physics, Massachusetts Institute of Technology, Cambridge, Massachusetts 02139, USA}
\author{Zhong Wang}
\affiliation{Institute for Advanced Study, Tsinghua University, Beijing 100084, China}
\affiliation{Collaborative Innovation Center of Quantum Matter, Beijing 100871, China}
\author{Chen Fang}
\affiliation{Institute of Physics, Chinese Academy of Sciences/Beijing National Laboratory for Condensed Matter Physics, Beijing 100190, China}
\affiliation{Department of Physics, Massachusetts Institute of Technology, Cambridge, Massachusetts 02139, USA}
\author{John D. Joannopoulos}
\author{Marin Solja\v{c}i\'{c}}
\affiliation{Department of Physics, Massachusetts Institute of Technology, Cambridge, Massachusetts 02139, USA}
\author{Liang Fu}
\affiliation{Department of Physics, Massachusetts Institute of Technology, Cambridge, Massachusetts 02139, USA}
\author{Nicholas X. Fang}\email{nicfang@mit.edu}
\affiliation{Department of Mechanical Engineering, Massachusetts Institute of Technology, Cambridge, Massachusetts 02139, USA}

\date{\today}

\begin{abstract}
Classical wave fields are real-valued, ensuring the wave states at opposite frequencies and momenta to be inherently identical. Such a particle-hole symmetry can open up new possibilities for topological phenomena in classical systems. Here we show that the historically studied two-dimensional (2D) magnetoplasmon, which bears gapped bulk states and gapless one-way edge states near zero frequency, is topologically analogous to the 2D topological $p+\Ii p$ superconductor with chiral Majorana edge states and zero modes. We further predict a new type of one-way edge magnetoplasmon at the interface of opposite magnetic domains, and demonstrate the existence of zero-frequency modes bounded at the peripheries of a hollow disk. These findings can be readily verified in experiment, and can greatly enrich the topological phases in bosonic and classical systems.
\end{abstract}

\maketitle

\pretolerance=8000 

Since the introduction of chiral edge states from two-dimensional (2D) quantum Hall systems into photonic systems \cite{Haldane:2008-PRL,Wang2009,hafezi2011robust,fang2012realizing,khanikaev2013photonic,rechtsman2013photonic,lu2014topological,zhang2015observation}, the investigation of band topology has been actively extended into many other 2D or 3D bosonic systems \cite{lu2015experimental,lu2015three}, including phonon \cite{prodan2009topological,kane2013topological,paulose2015topological,he2015acoustic,huber2016topological}, magnon \cite{shindou2013topological}, exciton \cite{yuen2014topologically,yuen2016plexciton}, and polariton \cite{karzig2015topological}. Up to now, topological phases of 2D plasmon have not been identified, in spite of some related discussion \cite{ling2015topological,cheng2015topologically,sinev2015mapping,yu2008one,gao2015topological,song2016chiral,kumar2015chiral} and a recent proposal of Weyl plasmon \cite{gao2015plasmon}. Besides, the previous works mainly focus on finite frequencies. Little attention has been paid to near-zero frequencies, where new symmetries and new topological states may emerge (as elaborated below).

Plasmon is a unique type of bosonic excitations. Microscopically, it consists of collective motion of electron-hole pairs in a Coulomb-interaction electron gas; whereas macroscopically, it appears as coherent electron-density oscillations. Under most circumstances, it can be well described by classical density (and velocity) fields on the hydrodynamic level. These fields, like all other classical fields, are intrinsically real-valued and respect an unbreakable particle-hole symmetry. For a particle-hole symmetric system, its Hamiltonian $\mathcal{H}$  transforms under an antiunitary particle-hole conjugate operation $\mathcal{C}$ via $\mathcal{C}^{-1}\mathcal{H}\mathcal{C}=-\mathcal{H}$. (Here $\mathcal{C}^2=+1$ for bosons.) This property ensures a symmetric spectrum $\omega(\mathbf{q})$ with respect to zero frequency, $\omega(\mathbf{q})=-\omega(-\mathbf{q})$, in which $\mathbf{q}$ is the wavevector. The associated wave fields are thus superpositions of complex-conjugate pairs, $\mathcal{F}(\mathbf{r},t) = \int\Dd\mathbf{q} \left[ \mathcal{F}_{\mathbf{q}} e^{+\Ii(\mathbf{q}\cdot \mathbf{r}-|\omega(\mathbf{q})| t)} + \mathcal{F}_{\mathbf{q}}^* e^{-\Ii(\mathbf{q}\cdot \mathbf{r}-|\omega(\mathbf{q})| t)}\right]$. The spin-0 real-scalar field governed by the Klein-Gordon equation, and the spin-1 real-vector field governed by Maxwell's equations also share the same feature.

Particle-hole symmetry greatly expands the classification of topological phases, according to the results of ten-fold classification \cite{Kitaev2009,Schnyder2008}. For example, the 2D quantum Hall phase belongs to the class-A in 2D with broken time-reversal symmetry $\mathcal{T}$. The Su-Schrieffer-Heeger~(SSH) model and the recently studied phonon zero modes \cite{kane2013topological,huber2016topological} belong to the class-BDI in 1D with particle-hole symmetry $\mathcal{C}^2=+1$ and time-reversal symmetry $\mathcal{T}^2=+1$.
However, the class-D 2D topological phase with $\mathcal{C}^2=+1$ and broken $\mathcal{T}$ has so far only been proposed in $p+\Ii p$ superconductors \cite{read2000paired}, which carry chiral Majorana edge states and Majorana zero modes.

In this work, we show that the historically studied 2D magnetoplasmon (MP) \cite{Stern1967PRL,Chiu1974PRB,Grimes1976PRL,Allen1977PRL,Theis1977SSC,glattli1985dynamical,Mast1985PRL,ando1982electronic,andrei1988low,Ashoori1992PRB,Aleiner1994PRL} belongs to the class-D 2D topological phase with unbreakable $\mathcal{C}$ and broken $\mathcal{T}$. It is governed by three-component linear equations carrying a similar structure as that of the two-band Bogoliubov-de Gennes (BdG) equations of the $p+\Ii p$ topological superconductor \cite{read2000paired,BernevigBook}. It contains a gapped bulk spectrum around zero frequency, and possesses gapless topological edge states and zero-frequency bound states (zero modes). Many properties of $p+\Ii p$ superconductor can find their analogy in 2D MP. Therefore, 2D MP provides the first realized example of class-D in the ten-class table \cite{Kitaev2009,Schnyder2008}. The experimentally observed edge MP states dated back to 1985 \cite{Mast1985PRL} are in fact topologically protected, analogous to the 1D chiral Majorana edge states. Some simulations of Majorana-like states in photonics have been reported \cite{tan2014photonic,rodriguez2014optical,keil2015optical,poddubny2014topological,iadecola2016non}, but they only appear at finite frequencies around which the particle-hole symmetry does not rigorously hold, unless using high nonlinearity\cite{peano2016topological}. Additionally, we are able to derive nonzero Chern numbers adhering to the band topology of 2D MP. Based on this, we are able to predict new type of one-way edge MP states on the boundary between opposite magnetic domains, and the existence of MP zero modes on a hollow disk. Our prediction can be experimentally verified in any two-dimensional electron gas (2DEG) systems, such as charged liquid-helium surface \cite{Grimes1976PRL}, semiconductor junctions \cite{Allen1977PRL,ando1982electronic,stone2007photovoltaic}, and graphene \cite{jablan2009plasmonics,KoppensNanoLett2011}.

\begin{figure*}[t]
\centerline{\includegraphics[scale=0.7]{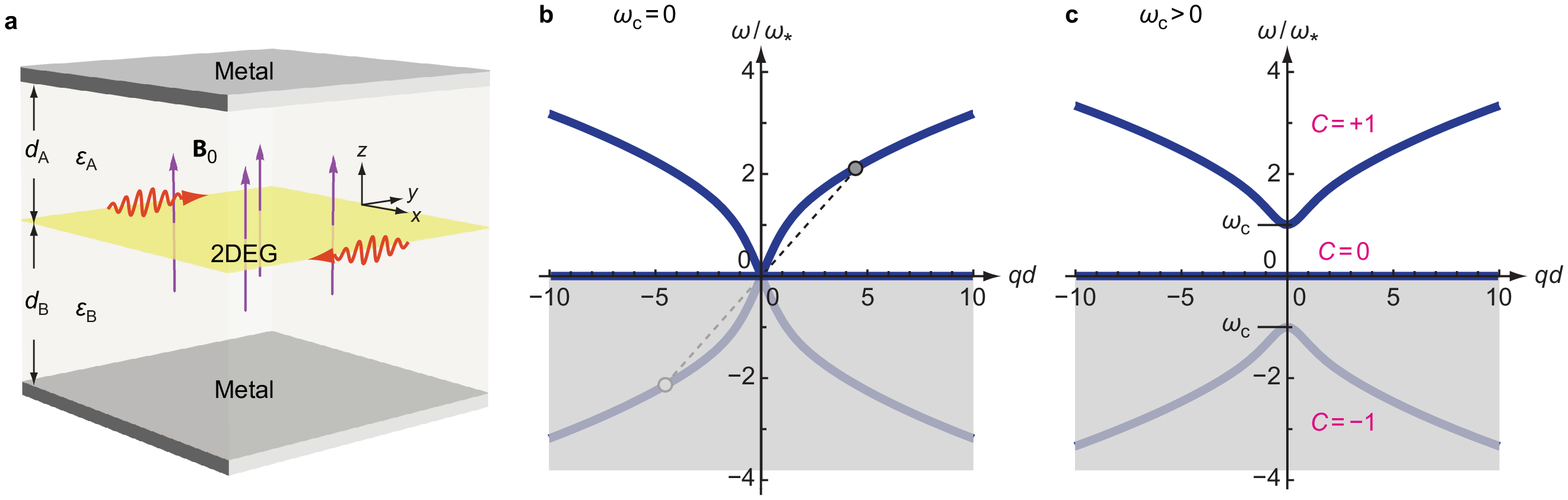}}\caption{Model system and analytically calculated homogeneous bulk spectra. (\textbf{a}) Schematics of the structure. (The presence of metal plates and dielectrics permits a general theoretical treatment but is not essential to the topological behaviors of 2D MP.) (\textbf{b}) No magnetic field is present. The two circles connected by a dashed line represent the particle-hole symmetry of the dispersion curves. (\textbf{c}) A uniform magnetic field is applied along the positive $z$-direction. Different branches carry different Chern numbers. The negative-frequency branch reflects the redundant degrees of freedom of the real-valued classical field and so is shaded. Here $\OmegaC=\omega_\star$, where $\omega_\star$ is a characteristic frequency defined to normalize the frequencies (see Methods: Theoretical model).}
\label{FigNonretardedBulk}
\end{figure*}

\ \\\noindent\textsf{\textbf{Results}}

\ \\\noindent\textbf{Governing equations.} We consider a 2DEG confined in the $z=0$ and $\mathbf{r}=x\Ex+y\Ey$ plane. The dynamics of MPs are governed by the linearized charge-continuity equation and the Lorentz-force equation \cite{Fetter1985PRB,Fetter1986PRB}. In the frequency domain, they are
\begin{align}
-\Ii\omega\rho(\mathbf{r},\omega) &= -\Nabla\cdot\mathbf{\calJ}(\mathbf{r},\omega), \label{EqnContinuity}\\
-\Ii\omega\mathbf{\calJ}(\mathbf{r},\omega) &= \mathit{\alpha}(\mathbf{r}) \mathbf{E}(\mathbf{r},\omega) - \OmegaC(\mathbf{r}) \mathbf{\calJ}(\mathbf{r},\omega)\times \Ez, \label{EqnEOM}
\end{align}
where
$\rho(\mathbf{r},\omega)$ is the variation of 2D electron density off equilibrium and $\mathbf{\calJ}(\mathbf{r},\omega)$ is the induced 2D current density. $\alpha(\mathbf{r})$ is a space-dependent coefficient that gives a 2D local longitudinal conductivity, $\sigma(\mathbf{r},\omega)=\Ii \alpha(\mathbf{r})/ \omega$. $\OmegaC(\mathbf{r})$ is the cyclotron frequency from a perpendicularly applied static magnetic field, $\mathbf{B}_0(\mathbf{r})=B_0(\mathbf{r})\Ez$. For the massive electrons chosen for our demonstration in this paper, $\alpha(\mathbf{r}) = \frac{e^2n_0(\mathbf{r})}{m_*}$ and $\OmegaC(\mathbf{r}) = \frac{eB_0(\mathbf{r})}{m_*c}$, in which $n_0(\mathbf{r})$ is the equilibrium electron density distribution, $m_*$ is the effective mass, and $c$ is the speed of light \cite{Chiu1974PRB,Volkov1988JETP}.

$\mathbf{E}(\mathbf{r},\omega)$ in Eq.~(\ref{EqnEOM}) is the electric field evaluated within the $z=0$ plane. It is generally a nonlocal function of the 2D current density, $\mathbf{E}(\mathbf{r},\omega)=\int \Dd\mathbf{r}' \ \mathbf{K}(\mathbf{r},\mathbf{r'},\omega)\cdot\mathbf{\calJ}(\mathbf{r}',\omega)$, where $\mathbf{K}(\mathbf{r},\mathbf{r'},\omega)$ is an integration kernel determined by Maxwell's equations. For a model system shown in Fig.~\ref{FigNonretardedBulk}a, and in the nonretarded limit ($\mathbf{E}$ solely comes from the Coulomb interaction) \cite{Fetter1985PRB,Fetter1986PRB}, the Fourier transformed field-current relation can be attested to be (see Supplementary Note 1)
\begin{equation}
\begin{split}
\begin{pmatrix}
E_x(\mathbf{q},\omega) \\
E_y(\mathbf{q},\omega)
\end{pmatrix}
=
\frac{2\pi}{\Ii\omega q \xi(q)}
\begin{pmatrix}
\DS q_x^2 & q_xq_y  \\
q_xq_y & \DS q_y^2
\end{pmatrix}
\begin{pmatrix}
\calJ_x (\mathbf{q},\omega)\\
\calJ_y (\mathbf{q},\omega)
\end{pmatrix}
.
\end{split}\label{EqnNonretardedField}
\end{equation}
Here $q=|\mathbf{q}|$, and
$\xi(q) = \frac{1}{2}\left\{\EpsA \coth(q\DdA)+ \EpsB \coth(q\DdB) \right\}$
is a $q$-dependent screening function \cite{Volkov1988JETP}.
A generalization of Eq.~(\ref{EqnNonretardedField}) can be made to include photon retardation (plasmon becomes plasmon-polariton) \cite{Chiu1974PRB,Kukushkin2003PRL,Deng2015PRB} (see Supplementary Note 2). The Drude loss by electron-phonon collision can be included in this formulation by introducing a finite lifetime in Eq.~(\ref{EqnEOM}) \cite{Volkov1988JETP}.

\ \\\noindent\textbf{Bulk Hamiltonian and bulk states.} We analytically solve for the case of homogeneous bulk plasmons, where $n_0(\mathbf{r})$, $B_0(\mathbf{r})$, $\alpha(\mathbf{r})$ and $\OmegaC(\mathbf{r})$ are all constants. $B_0$ and $\OmegaC$ can be positive or negative, depending on the direction of the magnetic field to be parallel or antiparallel to $\Ez$. We find that Eqs.~(\ref{EqnContinuity}--\ref{EqnNonretardedField}) can be casted into a Hermitian eigenvalue problem. The Hamiltonian $\mathcal{H}$ is block-diagonalized in the momentum space, acting on a generalized current density vector $\mathbf{J}$,
\begin{equation}
\omega \mathbf{J}(\mathbf{q},\omega) = \mathcal{H}(\mathbf{q}) \mathbf{J}(\mathbf{q},\omega) ,\quad \mathbf{J}(\mathbf{q},\omega)\equiv
\begin{pmatrix}
\JR(\mathbf{q},\omega)\\
\JD(\mathbf{q},\omega)\\
\JL(\mathbf{q},\omega)
\end{pmatrix},\label{EqnMatrixForm}
\end{equation}
where $\JLR(\mathbf{q},\omega) \equiv \frac{1}{\sqrt{2}}\{\calJ_x(\mathbf{q},\omega)\pm\Ii \calJ_y(\mathbf{q},\omega)\}$ are the left- and right-handed chiral components of current density, and
$\JD(\mathbf{q},\omega)\equiv \sqrt\frac{2\pi\alpha}{q\xi(q)}\rho(\mathbf{q},\omega)\equiv \frac{\OmegaP(q)}{q}\rho(\mathbf{q},\omega)$ is a generalized density component.
\begin{equation}
\OmegaP(q)=\sqrt{\frac{2\pi\alpha q}{\xi(q)}} \rightarrow
\begin{cases}
\vP q, \quad  & (\text{for small }q),\\
\sqrt{\frac{4\pi \alpha}{\EpsA+\EpsB}q}, \quad &(\text{for large }q),
\end{cases}
\end{equation}
is the dispersion relation of conventional 2D bulk plasmon without a magnetic field. $\vP = \sqrt{\frac{4\pi \alpha \DdA\DdB}{\EpsA\DdB+\EpsB\DdA}}$ is the effective plasmon velocity originated from the screening of Coulomb interaction at long wavelengths by surrounding metals (see Fig.~\ref{FigNonretardedBulk}a). $\JD(\mathbf{q},\omega)$ asymptotically equals $\vP\rho(\mathbf{q},\omega)$ at long wavelengths.

The bulk Hamiltonian $\mathcal{H}(\mathbf{q})$ and its long-wavelength limit read
\begin{widetext}
\begin{equation}
\mathcal{H}(\mathbf{q})
=
\begin{pmatrix}
+\OmegaC & \DS\frac{\OmegaP(q)}{q} \frac{q_x-\Ii q_y}{\sqrt{2}} & 0 \\
\DS\frac{\OmegaP(q)}{q} \frac{q_x+\Ii q_y}{\sqrt{2}} & 0 & \DS\frac{\OmegaP(q)}{q} \frac{q_x-\Ii q_y}{\sqrt{2}} \\
0 & \DS\frac{\OmegaP(q)}{q} \frac{q_x+\Ii q_y}{\sqrt{2}} & -\OmegaC
\end{pmatrix}
\overset{q\rightarrow0}{\longrightarrow}
\begin{pmatrix}
+\OmegaC & \DS\frac{\vP (q_x-\Ii q_y)}{\sqrt{2}} & 0 \\
\DS\frac{\vP (q_x+\Ii q_y)}{\sqrt{2}} & 0 & \DS\frac{\vP (q_x-\Ii q_y)}{\sqrt{2}} \\
0 & \DS\frac{\vP (q_x+\Ii q_y)}{\sqrt{2}} & -\OmegaC
\end{pmatrix}.
\label{EqnHamiltonian}
\end{equation}
\end{widetext}
Strikingly, this long-wavelength three-band $\mathcal{H}(\mathbf{q})$ has a very similar structure to that of the BdG hamiltonian of a 2D $p+\Ii p$ topological superconductor \cite{BernevigBook} which has two bulk bands.
Here, the odd number of bands is crucial for the topological consequences shown below.
For an even number of bosonic bulk bands, it has been proved that the summed Chern number is always zero for the bands below or above the zero frequency \cite{shindou2013topological}.

We can define the antiunitary particle-hole operation $\mathcal{C}$ and the antiunitary time-reversal operation $\mathcal{T}$ here,
\begin{equation}
\mathcal{C}=
\left.\begin{pmatrix}
0&0&1\\
0&1&0 \\
1&0&0
\end{pmatrix}
K \right|_{\mathbf{q}\rightarrow-\mathbf{q}},\
\mathcal{T}=
\left.\begin{pmatrix}
0&0&-1\\
0&1&0 \\
-1&0&0
\end{pmatrix}
K \right|_{\mathbf{q}\rightarrow-\mathbf{q}}
\end{equation}
where $K$ is the complex-conjugate operator and $|_{\mathbf{q}\rightarrow-\mathbf{q}}$ stands for a momentum flipping. As can be checked, $\mathcal{C}$ acts as the complex conjugation for all the field components $j_x$, $j_y$ and $\JD$, while $\mathcal{T}$ additionally reverses the sign for $j_x$ and $j_y$ (refer to Methods: Cartesian representation).
Our Hamiltonian has $\mathcal{C}$-symmetry,
\begin{equation}
\mathcal{C}\mathcal{H}(\mathbf{q})\mathcal{C}^{-1}=-\mathcal{H}(-\mathbf{q}),
\end{equation}
but broken $\mathcal{T}$-symmetry by the nonzero $\OmegaC$.

\begin{figure*}[t]
\centerline{\includegraphics[scale=0.7]{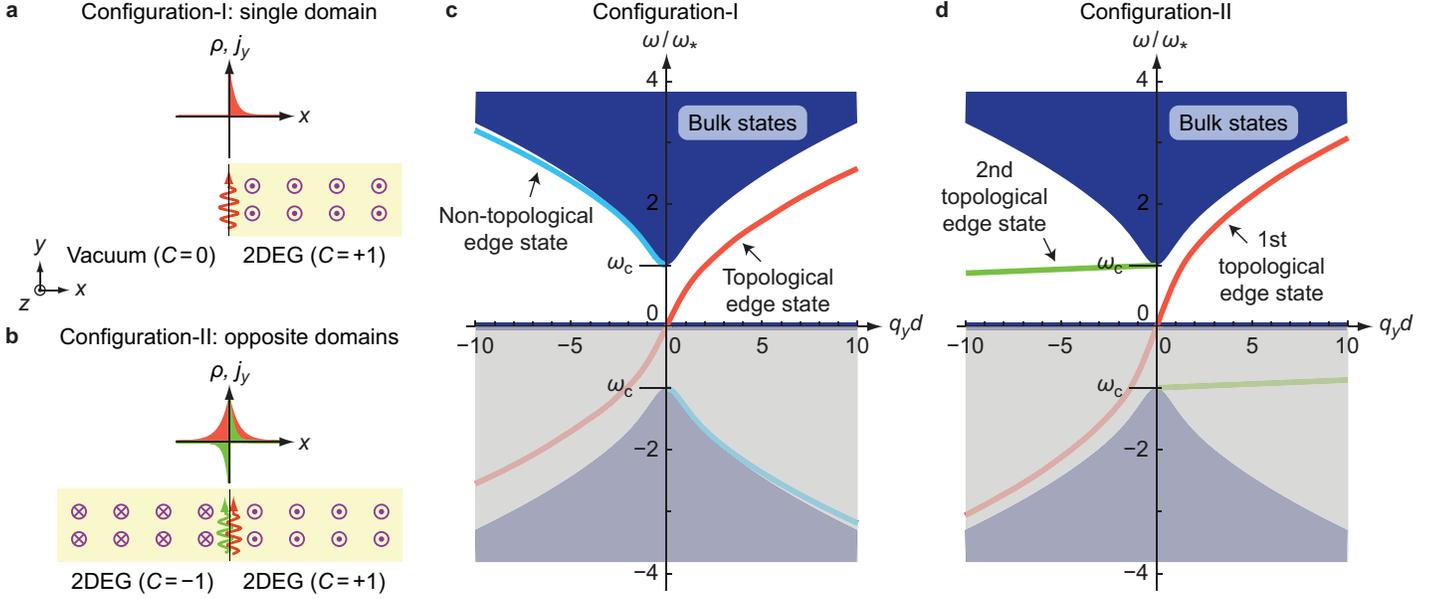}}\caption{
Schematics of the system configurations and numerically calculated bulk and edge spectra. (\textbf{a}) Configuration-I: A 2DEG is in contact with vacuum under a uniform magnetic field along the positive $z$-direction. (\textbf{b}) Configuration-II: A 2DEG is uniformly filled in the whole space but two opposite magnetic fields are applied along the positive and negative $z$-direction. The edge-state profiles of density and current components are shown as well. For Configuration-II, the two edge states have opposite symmetries. (\textbf{c}) The spectra corresponding to Configuration-I. (\textbf{d}) The spectra corresponding to Configuration-II. The negative-frequency part reflects the redundant degrees of freedom of the real-valued classical field and so is shaded.}
\label{FigNonretardedEdge}
\end{figure*}

The calculated homogeneous bulk spectra, plotted in Fig.~\ref{FigNonretardedBulk}b and c, are
\begin{equation}
\omega_{\SSS\pm}(\mathbf{q}) = \pm\sqrt{\OmegaC^2 + \OmegaP^2(q)} \quad \text{and}\quad \omega_{0}(\mathbf{q}) = 0.\label{EqnBulkSpectrum}
\end{equation}
The corresponding (unnormalized) wavefunctions are
\begin{align}
\mathbf{J}_{\SSS\pm}(\mathbf{q})\propto&
\begin{pmatrix}
  \DS\frac{q_x-\Ii q_y}{\sqrt{2}} \left(\sqrt{\OmegaC^2 + \OmegaP^2(q)} \pm\OmegaC\right) \\
  \pm \sqrt{\left(q_x^2+q_y^2\right)\OmegaP^2(q)} \\
 \DS\frac{q_x+\Ii q_y}{\sqrt{2}} \left(\sqrt{\OmegaC^2 + \OmegaP^2(q)} \mp\OmegaC\right)
\end{pmatrix}
\overset{q\rightarrow0}{\propto}
\begin{pmatrix}
\frac{1}{2}\pm\frac{1}{2}\\
0\\
\frac{1}{2}\mp\frac{1}{2}
\end{pmatrix}, \label{EqnPositiveBand}\\
\mathbf{J}_{0}(\mathbf{q})\propto&
\begin{pmatrix}
 \DS-\frac{q_x-\Ii q_y}{\sqrt{2}} \sqrt{\OmegaP^2(q)} \\
 \sqrt{q_x^2+q_y^2}\ \OmegaC \\
 \DS+\frac{q_x+\Ii q_y}{\sqrt{2}} \sqrt{\OmegaP^2(q)}
\end{pmatrix}
\overset{q\rightarrow0}{\propto}
\begin{pmatrix}0\\
1\\
0
\end{pmatrix}.
\label{EqnZeroBand}
\end{align}
The positive- and negative-frequency bands are gapped by $\OmegaC$.
The zero-frequency band represents purely rotational currents ($\Nabla\cdot\mathbf{\calJ}=0$, $\Nabla\times\mathbf{\calJ}\neq0$) balanced by static charge density distribution.

It is worthwhile to point out that the 2D bulk plasmon discussed here is distinctively different from the surface plasmon (or surface plasmon polariton) in a 3D boundary (see Supplementary Note 3).

\begin{figure}[htb]
\centerline{\includegraphics[scale=0.7]{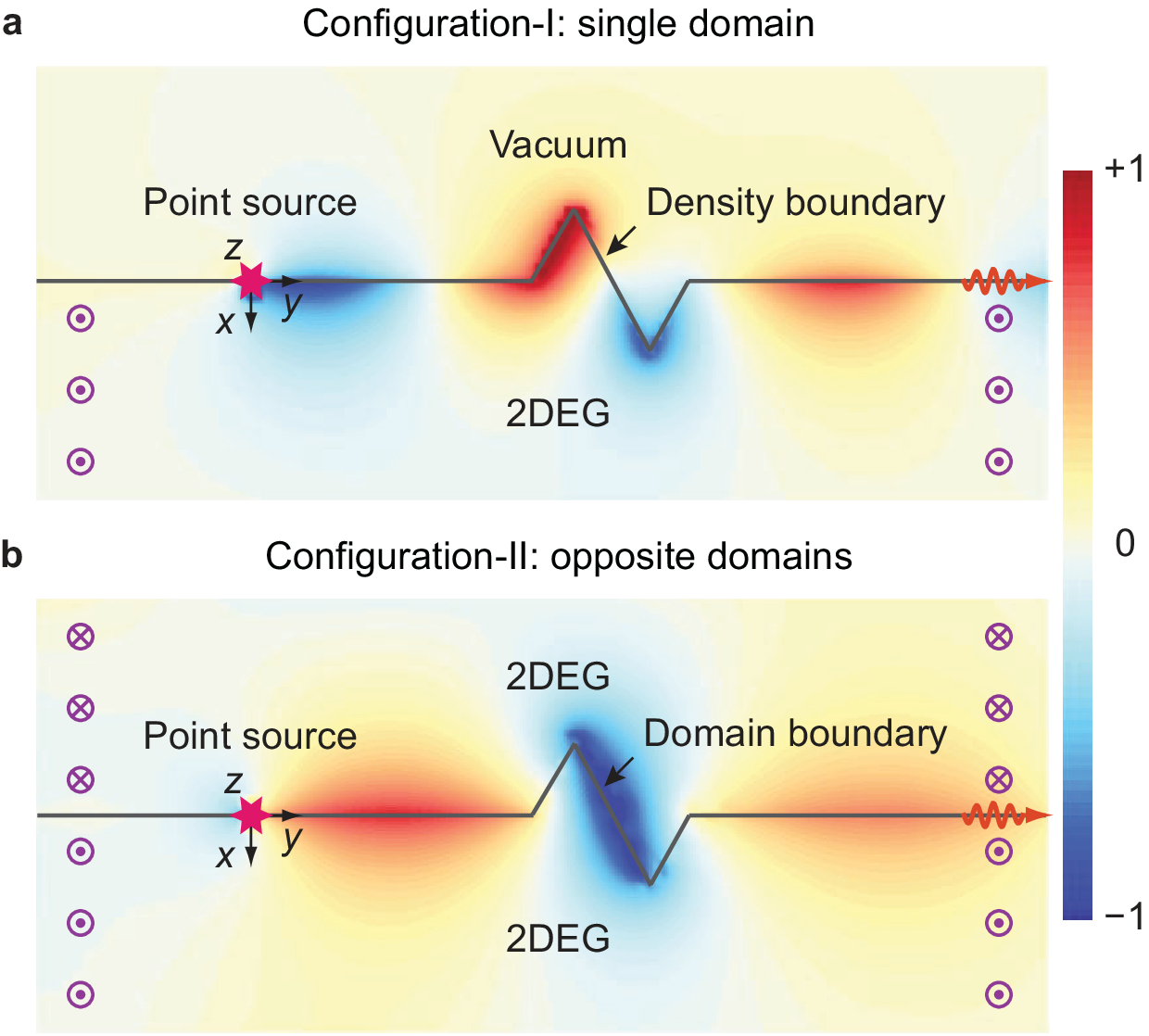}}\caption{Snapshots of one-way propagating topological edge states immune to backscattering at a sharp zigzag defect. Here the cyclotron frequency $\OmegaC=4\omega_*$, which corresponds to $B_0=0.91$~T . The driving frequency for the point source is $\frac{1}{2}\OmegaC$, which corresponds to $f=12.7$~GHz. (Refer to Methods: Theoretical model.) (\textbf{a}) Configuration-I: A 2DEG is in contact with vacuum under a single-domain magnetic field. (\textbf{b}) Configuration-II: A 2DEG is uniformly filled in the whole space under an opposite-domain magnetic field. Note that the profile is plotted for the electric scalar potential, which is nonzero in the vacuum region. The difference between (a) and (b) on the wavelengths is consistent with the slightly different dispersion slope of the topological edge modes between Configuration-I and II. (See online Supplementary Movies 1 and 2 for real-time evolution.)}\label{FigMovie}
\end{figure}

\ \\\noindent\textbf{Chern number on an infinite momentum plane.} Unlike the regular lattice geometries whose Brillouin zone is a torus, our system is invariant under continuous translation, where the unbounded wavevector plane can be mapped onto a Riemann sphere \cite{read2000paired,silveirinha2015chern}. As long as the Berry curvature decays faster than $q^{-2}$ as $q\rightarrow\infty$, the Berry phase around the north pole ($q=\infty$) of Riemann sphere is zero and the Chern number $C$ is quantized \cite{silveirinha2015chern}. For our 2D MP,
we can verify this analytically,
\begin{align}
C_{\SSS\pm} =&\ \frac{1}{2\pi} \int \Dd\mathbf{S}_{\mathbf{q}}\cdot\left\{\nabla_{\mathbf{q}} \times \frac{\left[\mathbf{J}_{\SSS\pm}(\mathbf{q}) \right]^\dagger (-\Ii\nabla_{\mathbf{q}})\mathbf{J}_{\SSS\pm}(\mathbf{q})}{\left[\mathbf{J}_{\SSS\pm}(\mathbf{q}) \right]^\dagger \mathbf{J}_{\SSS\pm}(\mathbf{q})} \right\} \nonumber\\
=&\int q\Dd q \left\{ \frac{1}{q} \partial_q \left[ \frac{\mp\OmegaC }{\DS \sqrt{\OmegaC^2 + \OmegaP^2(q)}} \right] \right\} \label{EqnChernNumber} \\
=&\ \pm \text{sgn}(\OmegaC), \nonumber \\
C_{0} =&\ 0,
\end{align}
where $\Dd\mathbf{S}_{\mathbf{q}}$ is the momentum-space surface element and the term in the curly bracket is the Berry curvature.
Since the Berry curvature changes its sign under the particle-hole symmetry, the positive and negative-frequency bands must have opposite Chern numbers, and the zero-frequency band, with odd Berry curvature, must have zero Chern number.

\begin{figure*}[t]
\centerline{\includegraphics[scale=0.7]{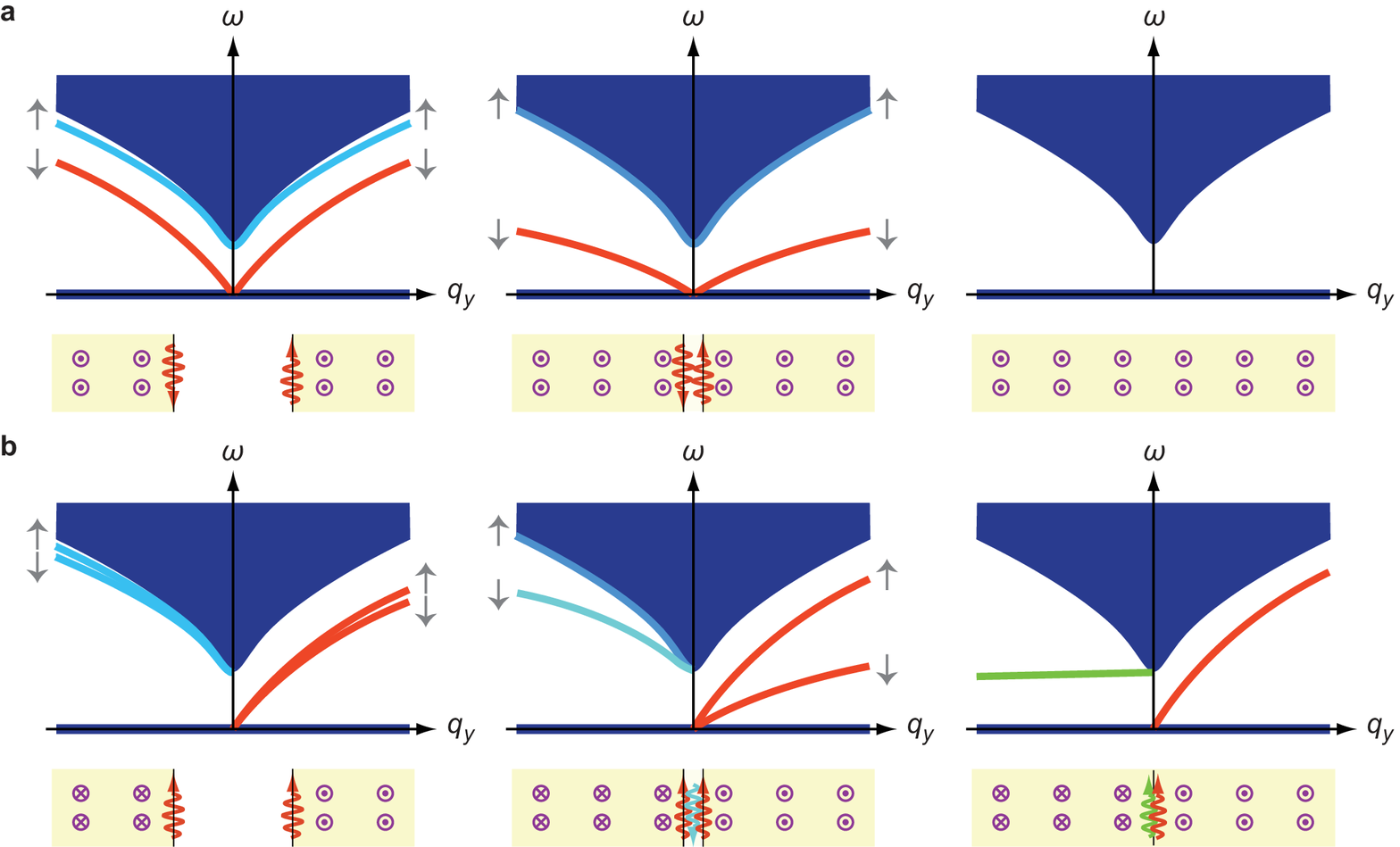}}
\caption{
Evolution of the edge states for two different configurations as the separated edge states on two pieces of 2DEG are brought into contact with each other. The density spacer is gradually filled to the bulk value as the distance shrinks to zero. (\textbf{a}) The magnetic field is uniformly applied along the positive $z$-direction in the whole space. (\textbf{b}) The magnetic field has opposite directions between the left-part and right-part of 2DEG.
}\label{FigEvolution}
\end{figure*}

\ \\\noindent\textbf{Majorana-type one-way edge states.} We are able to calculate the analytical edge solutions in the long-wavelength limit $q\rightarrow 0$, and numerical edge solutions for unrestricted wavelengths, as plotted in Fig.~\ref{FigNonretardedEdge}.
When $q\rightarrow 0$, the Hamiltonian in Eq.~(\ref{EqnHamiltonian}) becomes local, allowing us to replace $q_x$ and $q_y$ with the operators $-\Ii\partial_x$ and $-\Ii\partial_y$ and to solve for the edge states by matching boundary conditions \cite{Fetter1986PRB,song2016chiral}. We consider a 1D edge situated at $x=0$ caused by a discontinuity either in $n_0(x)$ (Configuration-I) or in $B_0(x)$ (Configuration-II).

In Configuration-I as shown in Fig.~\ref{FigNonretardedEdge}a, we let the $x>0$ region be filled with 2DEG, ($\vP(x>0)=\vP$), while the $x<0$ region be vacuum, ($\vP(x<0)=0$), and the magnetic field be uniformly applied parallel to $\Ez$, ($\OmegaC(x)=\OmegaC>0$). Since we know that the change of Chern number across the edge is $\Delta C=\pm1$, there must be a single topological edge state present in this configuration. We look for solutions that behave like $\sim\Ee^{-\kappa x+\Ii q_y y -\Ii\omega t}$, ($q_x=\Ii\kappa$), in the $x>0$ region, where $\kappa>0$ is an evanescent wavenumber.
The boundary condition is $
\calJ_x|_{x=0^+} = 0$,
meaning that the component of the current normal to the edge must vanish. So the edge solutions have to satisfy $
q_y \OmegaC = \kappa \sqrt{\OmegaC^2+ \vP^2 (q_y^2-\kappa^2)}$,
whose right-hand side is always positive. Since we have chosen $\OmegaC=|\OmegaC|>0$, $q_y$ must be positive too. The edge state is hence one-way propagating.
(We discard the unphysical solution $\kappa = q_y$ which yields a null wavefunction.)
The physical solution is $\kappa = \frac{\OmegaC}{\vP}$ which gives an edge-state spectrum,
\begin{equation}
\omega^{\SSS(1)}_{\SSS\text{edge}}(q_y) = \vP q_y,\quad
(q_y\geq0).
\end{equation}
This is a gapless state running across the zero-frequency band as shown in Fig.~\ref{FigNonretardedEdge}c. It is a bosonic analogue to the chiral Majorana edge states in the topological superconductor. Its wavefunction is
\begin{equation}
\mathbf{J}^{\SSS(1)}_{\SSS\text{edge}}(x,q_y) \propto \begin{pmatrix}-\frac{\Ii}{\sqrt{2}}\\1\\+\frac{\Ii}{\sqrt{2}}\end{pmatrix}e^{-\frac{\OmegaC}{\vP}x+\Ii q_y y}, \quad (x>0).
\end{equation}
The zero-frequency edge state at $q_y=0$ has the same finite decay length $\kappa=\frac{\OmegaC}{\vP}$ and is well localized at the edge, despite being degenerate with the flat middle band.

In Configuration-II as shown in Fig.~\ref{FigNonretardedEdge}b, we let the 2DEG have a constant electron density throughout the whole space, ($\vP(x)=\vP$), but the magnetic field have opposite signs in the two regions, ($\OmegaC(x)=+|\OmegaC|>0$ for $x>0$ and $\OmegaC(x)=-|\OmegaC|<0$ for $x<0$). This is a novel configuration with $\Delta C=\pm2$ across the edge, and permits two topological edge states. The boundary conditions are $\calJ_x|_{x=0^-} = \calJ_x|_{x=0^+}$ and $\JD|_{x=0^-} = \JD|_{x=0^+}$, meaning that both the normal current and density must be continuous across the edge.
In our long-wavelength approximation, the first edge solution has the same spectrum and wavefunction as those of Configuration-I, except for a symmetric extension of the wavefunction into the $x<0$ region (compare the plots of $\rho$ and $\calJ_y$ in red in Fig.~\ref{FigNonretardedEdge}a and b). The second edge solution satisfies $\kappa=-q_y$, and so $q_y$ must be negative to ensure $\kappa$ positive. This edge state is also one-way with a spectrum,
\begin{equation}
\omega^{\SSS(2)}_{\SSS\text{edge}}(q_y) =|\OmegaC|,\quad(q_y\leq0),
\end{equation}
as shown in Fig.~\ref{FigNonretardedEdge}d. Its wavefunction is
\begin{equation}
\mathbf{J}^{\SSS(2)}_{\SSS\text{edge}}(x,q_y) \propto
\begin{pmatrix}\frac{1}{2}\pm\frac{1}{2}\\0\\\frac{1}{2}\mp\frac{1}{2}\end{pmatrix}e^{\pm q_y x+\Ii q_y y}, \quad (x\gtrless0).
\end{equation}
It is antisymmetric for $\calJ_y$ as plotted in green in Fig.~\ref{FigNonretardedEdge}b. The antisymmetry plus the continuity condition for $\JD$ here make the density oscillations vanish identically, so the second edge state carries only current oscillations about the magnetic domain wall.

Figure \ref{FigNonretardedEdge}c and d give our numerically calculated bulk and edge spectra that are not restricted by the small-$q$ local approximation (refer to Methods: Numerical scheme). When $q$ is small, the numerical results accurately agree with our analytical derivation above. When $q$ is large, the 1st edge dispersion asymptotically approaches the $\sqrt{q}$-like bulk bands and eventually connects to them at $q\rightarrow\infty$.
The 2nd edge dispersion drops gradually towards the zero-frequency bulk band and connects to it at $q\rightarrow\infty$. It does show a (tiny) positive group velocity indicating the correct chiral direction, in spite of its negative phase velocity (refer to Fig.~\ref{FigNonretardedEdge}b and d). Physically though, the plasmon picture breaks down in the large-$q$ regime, where electron-hole pair production takes place.

Besides, we should note there exists a non-topological gapped edge state in Configuration-I (see the blue curve in Fig.~\ref{FigNonretardedEdge}c), which is consistent with the literature \cite{Volkov1988JETP}. At long wavelengths $q\rightarrow0$, this state resembles a bulk state bearing an excitation gap of $\OmegaC$. Its edge confinement shows up only at short wavelengths. As elaborated below, under a parameter evolution from Configuration-I to II, this edge state evolves, remarkably, into the 2nd gapless topological edge state in Configuration-II.

\ \\\noindent\textbf{One-way propagation of edge states.} Owing to the topological protection, one-way edge states are immune to backscattering from a random defect. In Fig.~\ref{FigMovie}, we display two snapshots from our real-time simulation. Edge states are excited by a point source oscillating at a frequency of $\frac{1}{2}\OmegaC$ within the bulk gap. The generated waves propagate only towards the right. A sharp zigzag defect has been purposely inserted into the route of propagation. The waves then exactly follow the edge profile and insist on propagating forward without undergoing any backscattering. The slightly visible fluctuation to the left of the point source is completely local (non-propagating). It is caused by the unique long-range Coulomb interaction in this system, different from the more familiar photonic and acoustic systems.

Fig.~\ref{FigMovie}a corresponds to the traditional Configuration-I, where the physical edge is formed by the density termination of 2DEG at vacuum. Although the one-way nature in this configuration is known historically, its absolute robustness due to the topological protection is less known and is manifested here.

Fig.~\ref{FigMovie}b corresponds to our new Configuration-II. Only the 1st edge state classified in Fig.~\ref{FigNonretardedEdge} is excited here; the 2nd edge state has an energy too close to the band edge $\OmegaC$ and hence is not excited here. Based on our argument on topology above, Configuration-II permits protected one-way edge states on a magnetic domain boundary even if the 2DEG may be homogeneous. Fig.~\ref{FigMovie}b vividly demonstrates this scenario. Furthermore, it shows the perfect immunization to backscattering when the two magnetic domains drastically penetrate each other.

\ \\\noindent\textbf{Evolution of edge states.} We find that the adiabatic mode evolution of the MP edge states described by a three-band model here is rather different from that in 2D topological insulators or superconductors, which can be commonly described by a two-band or four-band model. The zero-frequency bulk band here plays a critical role for the appearance and disappearance of the topological edge states. We have investigated the evolution corresponding to the aforementioned two configurations shown in Fig.~\ref{FigNonretardedEdge}.

We first study the interaction between two topological edge states propagating along the opposite directions on two 2DEG edges in the same magnetic field. The result is shown in Fig.~\ref{FigEvolution}a. When we gradually narrow the spacer, we let the equilibrium density in the spacer to gradually change from 0 to the bulk density $n_0$. This allows the two topological edge states to couple and end up with a bulk configuration. Instead of opening a gap and entering the top band (which would happen in the conventional two-band system), the two topological edge states here (marked in red) fall down into the flat middle band. In addition, there are two non-topological edge states (marked in blue, and refer to Fig.~\ref{FigNonretardedEdge}c) in this configuration. They move completely into the bulk during this process.

\begin{figure}[htb]
\centerline{\includegraphics[scale=0.7]{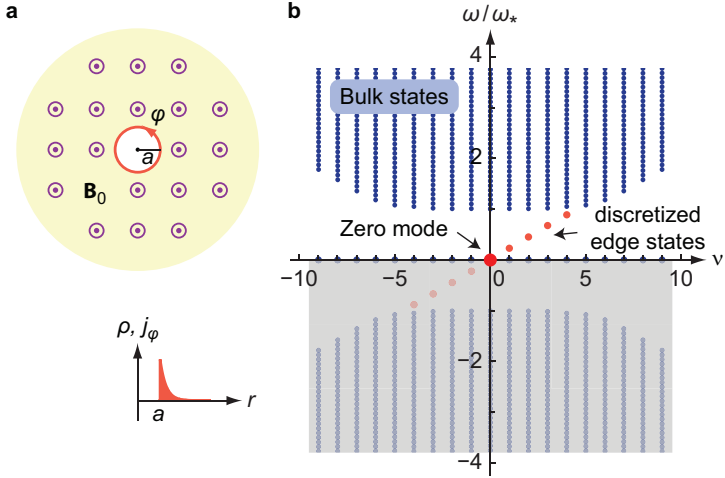}}\caption{Topological zero mode.
(\textbf{a}) Illustration of a hollow disk geometry and zero-mode profile attached to the inner circular edge. (\textbf{b}) Spectrum of bulk states and edge states versus the discrete angular momentum $\nu=0,\pm1,\pm2,\dots$.
}\label{FigZeromode}
\end{figure}

We then study the interaction between two topological edge states propagating along the same direction on the two 2DEG edges under opposite magnetic fields. The result is shown in Fig.~\ref{FigEvolution}b. When we gradually narrow the spacer, we find that one of the edge states (marked in red) goes slightly upward while the other falls downward into the flat middle band. In the meanwhile, one of the non-topological edge states (marked in blue) detaches from the top band, continuously bends downward until it reaches the frequency $\OmegaC$, where it transits into the second topological edge state (marked in green). According to our previous argument, this edge state connects to the middle band at the momentum infinity. At all times, there are two gapless edge states. This is consistent with the preserved difference of Chern numbers ($\Delta C=\pm2$) between the left-part and right-part of 2DEG.

\ \\\noindent\textbf{Magnetoplasmon on a hollow disk.} For infinitely long edges as discussed above, the edge states when $k_y\rightarrow0$ merge into the zero-frequency middle band, as can be seen in Fig.~\ref{FigNonretardedEdge}c and d. Importantly, we want to quest whether such zero-frequency modes preserve (not being gapped) when we wrap a long edges into small circle. We want to investigate the behaviors of 2D MP on, for instance, a hollow disk geometry depicted in Fig.~\ref{FigZeromode}a.

We write the low-energy long-wavelength Hamiltonian in real-space polar coordinates and still use the chiral representation $\JLR(r,\phi,\omega) \equiv \frac{1}{\sqrt{2}}\{\calJ_r(r,\phi,\omega)\pm\Ii \calJ_\phi(r,\phi,\omega)\}\Ee^{\pm\Ii\phi}$, $\JD(r,\phi,\omega) \equiv \vP\rho(r,\phi,\omega)$,
\begin{widetext}
\begin{equation}
\begin{pmatrix}
+\OmegaC & \DS \frac{\vP\Ee^{-\Ii\phi}}{\Ii \sqrt{2}}\left( \partial_r-\frac{\Ii}{r}\partial_\phi \right) & 0 \\
\DS \frac{\vP\Ee^{+\Ii\phi}}{\Ii \sqrt{2}}\left( \partial_r+\frac{\Ii}{r}\partial_\phi \right) & 0 & \DS \frac{\vP\Ee^{-\Ii\phi}}{\Ii \sqrt{2}}\left( \partial_r-\frac{\Ii}{r}\partial_\phi \right) \\
0 & \DS \frac{\vP\Ee^{+\Ii\phi}}{\Ii \sqrt{2}}\left( \partial_r+\frac{\Ii}{r}\partial_\phi \right) & -\omega_c
\end{pmatrix}
\begin{pmatrix}
\JR(r,\phi,\omega)\\
\JD(r,\phi,\omega)\\
\JL(r,\phi,\omega)
\end{pmatrix}
=
\omega
\begin{pmatrix}
\JR(r,\phi,\omega)\\
\JD(r,\phi,\omega)\\
\JL(r,\phi,\omega)
\end{pmatrix}. \label{EqnPolarHamiltonian}
\end{equation}
\end{widetext}
The azimuthal symmetry ensures the eigensolutions of $\calJ_r$, $\calJ_\phi$, and $\JD$ to pick up a common factor $\Ee^{\Ii\nu\phi}$, where $\nu$ is the integer of discretized angular momentum. This low-energy long-wavelength problem can be solved semi-analytically. In all situations, $\calJ_r$ and $\calJ_\phi$ are derived quantities from $\JD$,
\begin{align}
\calJ_r(r,\nu,\omega) &= \frac{-\Ii\vP}{\omega^2-\OmegaC^2} \left[ \OmegaC\frac{\nu}{r}+\omega\partial_r \right]\JD(r,\nu,\omega) , \label{EqnCircularJR} \\
\calJ_\phi(r,\nu,\omega) &= \frac{\vP}{\omega^2-\OmegaC^2} \left[ \omega \frac{\nu}{r}+\OmegaC \partial_r \right]\JD(r,\nu,\omega) . \label{EqnCircularJPhi}
\end{align}

Depending on the energy to be above or below the bandgap, we have
\begin{equation}
\JD(r,\nu,\omega) =
\begin{cases}
\mathcal{A}_\nu \mathrm{J}_{\nu} ( k_r r ) + \mathcal{B}_\nu \mathrm{Y}_{\nu} ( k_r r ),\quad (\omega^2<\OmegaC^2),\\
\mathcal{C}_\nu \mathrm{K}_\nu(\kappa_r r), \quad (0<\omega^2<\OmegaC^2).
\end{cases}
\label{EqnCircularJD}
\end{equation}
where $k_r=\sqrt{\omega^2-\OmegaC^2}/\vP$, $\kappa=\sqrt{\OmegaC^2-\omega^2}/\vP$, $\mathrm{J}_\nu(k_rr)$ and $\mathrm{Y}_\nu(k_rr)$ are the $\nu$-th order Bessel and Neumann functions, and $\mathrm{K}_\nu(\kappa_rr)$ is the modified Bessel function. $\mathcal{A}_\nu$, $\mathcal{B}_\nu$, and $\mathcal{C}_\nu$ are coefficients. The no-normal-current boundary condition $\calJ_r(r,\nu,\omega) |_{r=a}=0$ determines the allowed eigenfrequencies $\omega$ for every given $\nu$. In this approximate model, the solutions above the bandgap $\omega^2<\OmegaC^2$ are all continuous bulk-like and within the bandgap $0<\omega^2<\OmegaC^2$ are all discrete edge-like, as shown in Fig.~\ref{FigZeromode}b.

\ \\\noindent\textbf{Majorana-type bound state: zero mode.} We are particularly interested in the limit $\omega\rightarrow0$. According to Eqs.~(\ref{EqnCircularJR}) and (\ref{EqnCircularJD}), the boundary condition can be satisfied if and only if $\nu=0$, which completely annihilates the radial current, $\calJ_r(r,0,0)=0$. The zero-mode profile is given by
\begin{align}
\JD(r,0,0) & = \mathcal{A}_0 \mathrm{K}_{0} \left( \frac{|\OmegaC|}{\vP} r \right) \sim \frac{\Ee^{-\frac{|\OmegaC|}{\vP}r}}{\sqrt{\frac{|\OmegaC|}{\vP}r}},\\
\calJ_\phi(r,0,0) & = \mathcal{A}_0 \mathrm{sgn}(\OmegaC) \mathrm{K}_{1} \left( \frac{|\OmegaC|}{\vP} r \right) \sim \frac{\mathrm{sgn}(\OmegaC)\Ee^{-\frac{|\OmegaC|}{\vP}r}}{\sqrt{\frac{|\OmegaC|}{\vP}r}}. \nonumber
\end{align}
$\JD$ describes a static charge distribution; $\calJ_\phi$ describes a dc circulating current. The electric field generated by $\JD$ balances out the Lorentz force due to $\calJ_\phi$ and $\mathbf{B}_0$. The overall amplitude and phase in $\mathcal{A}_0$ can be arbitrary, but $\JD$ and $\calJ_\phi$ are phase-locked to each other by a ratio of $\mathrm{sgn}(\OmegaC)$, i.e., the sign of magnetic field. In the chiral representation, with the arbitrary phase ignored, the zero-mode wavefunction behaves like
\begin{equation}
\mathbf{J}_{\SSS\text{zero-mode}}(r) \propto
\begin{pmatrix}
-\frac{\Ii}{\sqrt{2}} \Ee^{-\Ii\phi} \\ \mathrm{sgn}(\OmegaC) \\ +\frac{\Ii}{\sqrt{2}} \Ee^{+\Ii\phi}
\end{pmatrix}
\frac{\Ee^{-\frac{|\OmegaC|}{\vP}r}}{\sqrt{\frac{|\OmegaC|}{\vP}r}}, \quad (r > a).\label{EqnZeroModeWavefunction}
\end{equation}

We see that the Majorana-type topological edge state within the bandgap indeed preserve as $\omega\rightarrow0$ and $\nu=0$, on the finite-sized inner edge of a hollow disk. It is easy to verify that shrinking the hole radius will only reduce the number of discrete edge states at a finite $\nu$, but not kill the zero mode. These observations are consistent with the $k_y\rightarrow0$ limit of the infinite straight edge case (see Fig.~\ref{FigNonretardedEdge}c and d). Nevertheless, we should cautiously note that the zero mode here is not isolated. It degenerates with a large number of other zero-frequency modes in the middle band (see Supplementary Note 4).

The guaranteed existence of zero mode in a hollow disk geometry for 2D MP is reminiscent of the Majorana zero modes in the 2-band $p+\Ii p$ topological superconductor \cite{read2000paired}. In the latter case a $\pi$ flux inside the vortex core is needed to balance the anti-periodic boundary condition in the $\phi$ direction. By comparison, our 3-band Hamiltonian here hosts zero mode with periodic boundary condition, i.e. the wavefunction in Eq.~(\ref{EqnZeroModeWavefunction}) returns to itself when $\phi$ rotates $2\pi$.

More generally, similar kind of discretized edge states and zero modes must exist on a noncircular geometry as well, and can be on both the inner and outer edges, if the geometry is finite. For a finite geometry, the inner and outer edge zero modes must come out in pairs. If one closes the inner boundary and turns the hollow disk into a solid disk, then both modes must vanish. Otherwise, the vortex currents associated with both zero modes are singular at the center, and are unphysical.

\ \\\noindent\textsf{\textbf{Discussion}}
\ \\

\noindent Although the configuration of 2D MP looks almost identical to that of the quantum Hall effect (QHE), there are fundamental differences \cite{andrei1988low}. QHE deals with electron (fermionic) transport, whereas 2D MP deals with collective electron-density oscillations (bosonic excitations). In QHE, the bulk is electronically insulating, whereas in 2D MP, the bulk is electronically conducting. QHE experiments usually require a high magnetic field ($\gtrsim8$~T) and a high electron density ($\gtrsim10^{11}$~cm$^{-2}$). By comparison, MP experiments normally does not require a high magnetic field or a high electron density. Moreover, the bandgap in QHE is for electron transport, while the bandgap for 2D MP is for electron-density oscillations.

2D MP belongs to the topological class-D, which is different from the quantum Hall class-A. The topological edge states in MP are not governed by the traditional bulk-edge correspondence of the quantum Hall states. The traditional rule states that the number of gapless edge states inside a gap is equal to the sum of all the bulk Chern numbers from the energy zero up to the gap. In our topological MP, the lowest bulk band (zero-frequency flat band) contributes zero Chern number.
A generalized bulk-edge correspondence must be established by considering the particle-hole symmetry that extends the spectrum into the ``redundant" negative-frequency regime. This accounts for the Berry flux exchanged across the zero frequency. Only by doing so, the sum of all the bulk Chern numbers, spanning both the positive and negative-frequency regimes, can be correctly kept zero.

Experiments for 2D plasmon were performed by pioneering researchers dated back to 1970s on the charged liquid helium surface \cite{Grimes1976PRL} and in the semiconductor inversion layer \cite{Allen1977PRL}. 2D magnetoplasmon was first observed in the semiconductor inversion layer in 1977 \cite{Theis1977SSC} and edge magnetoplasmon was first observed on the liquid helium surface in 1985 \cite{Mast1985PRL} by measuring radio-frequency absorption peaks. The nonreciprocity of edge MP was verified in semiconductor heterojunctions at microwave frequencies \cite{Ashoori1992PRB}. Recently, similar experiments have been performed on 2D materials \cite{Crassee2012NL,yan2012infrared,Petkovic2013PRL,kumada2014resonant} and on the surface of a topological insulator \cite{autore2015observation}. However, all the previous studies on edge MPs correspond to Configuration-I in our paper, where the physical edge is formed between a 2DEG and vacuum under a uniform magnetic field.

\begin{figure}[htb]
\centerline{\includegraphics[scale=0.7]{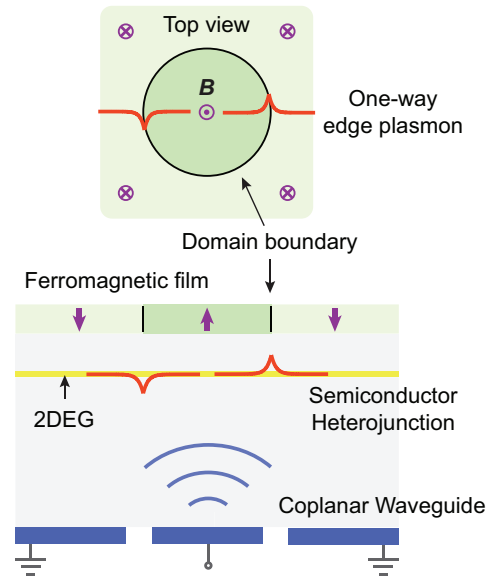}}\caption{Schematics of a proposed experimental design based on the 2DEG in a semiconductor heterojunction. Microwave transmitted on the coplanar waveguide can excite one-way edge plasmons at the boundary of opposite magnetic domains. Characteristic resonant absorption can be observed.}\label{FigExperiment}
\end{figure}

Our prediction of new one-way edge magnetoplasmon in Configuration-II has not been reported, despite there have been similar studies in the QHE systems \cite{Reijniers2000JPCM,Ye1995PRL}. Our prediction can in principle be verified in any 2DEG system as long as a pair of opposite magnetic fields is applied across. The magnetic domain boundary can be experimentally created by, for example, two concentric solenoids of opposite currents or ferromagnetic materials. The gapless one-way edge modes can be mapped out by microwave bulk transmission and by Fourier transforming near-field scans along the edges, as demonstrated in Ref.~\cite{skirlo2015experimental} for a topological photonic system. In Fig.~\ref{FigExperiment}, we provide a schematic experimental design. A semiconductor heterojunction serves as the platform for 2DEG. A ferromagnetic film is grown on the top and is polarized into up- and down-domains. For the circular domain boundary as sketched, discretized one-way edge states can travel around. A meandering coplanar waveguide can be fabricated on the back of the substrate. One can then send in and receive microwave signals at a frequency below $\OmegaC$, and in the meanwhile monitor the characteristic absorption, which signifies the excitation of one-way edge plasmons.

Our proposed zero-frequency bound modes feature localized charge accumulation (inducing a static electric field) and circulating current (inducing a static magnetic field). These effects are in principle measurable when the zero mode is actuated by charge or current injection.

In summary, we have revealed the salient topological nature of 2D MP as the first realization of the class-D topological phase. The predicted new Majorana-type one-way edge states and zero modes can be verified experimentally. Our work opens a new dimension for topological bosons via introducing the bosonic phase protected by the particle-hole symmetry. We anticipate similar realizations for other bosonic particles \cite{chen2015symmetry} and discoveries of new topological phases, after combining the particle-hole symmetry with other symmetries \cite{chiu2015classification}, for example, the time-reversal and many kinds of spatial symmetries \cite{fu2011topological,lu2015three}.

\ \\\noindent\textsf{\textsf{\textbf{Methods}}}
{
\small

\ \\\noindent\textsf{\textbf{Theoretical model.}} Our theoretical model of 2D MP is based on a hydrodynamic formalism. In equilibrium, the 2D electron number density is $n_{0}(\mathbf{r})$, which is determined by the device structure, doping concentration, and gating condition \cite{ando1982electronic}. Off equilibrium, the density changes to $n(\mathbf{r},t)$. For 2D MP, we are only concerned with the small deviation of 2D charge density,
$
\rho(\mathbf{r},t) = -e\left\{n(\mathbf{r},t) - n_{0}(\mathbf{r})\right\}
$,
and the induced 2D current density up to the linear order,
$
\mathbf{\calJ}(\mathbf{r},t)=-en_0(\mathbf{r})\mathbf{v}(\mathbf{r},t),
$
where $\mathbf{v}(\mathbf{r},t)$ is the local velocity of electron gas restricted to move in the $z=0$ plane only. For massive electrons, the conductivity coefficient is $\alpha(\mathbf{r}) = \frac{e^2n_0(\mathbf{r})}{m_*}$ and the cyclotron frequency is $\OmegaC(\mathbf{r}) = \frac{eB_0(\mathbf{r})}{m_*c}$ \cite{Chiu1974PRB,Volkov1988JETP}.
For massless electrons in, for example, graphene, they are generalized to $\alpha(\mathbf{r}) = \frac{e^2v_{\SSS\text{F}}\sqrt{n_0(\mathbf{r})}}{\hbar\sqrt{\pi}}$ and $\OmegaC(\mathbf{r}) = \frac{e v_{\SSS\text{F}}^2B_0(\mathbf{r})}{c\EF}$, where $v_{\SSS\text{F}}$ is the Fermi velocity and $\EF$ is the Fermi energy \cite{Wang2011PRB,Sounas2011APL}. Hence our theory works for both massive and massless electrons.

We consider a model system shown in Fig.~\ref{FigNonretardedBulk}(a). With different choices of the material constants and structural parameters, it can lead to different practical 2DEG systems, such as charged liquid-helium surface, metal-insulator-semiconductor junction, top-gated graphene transistor, etc. In Fig.~\ref{FigNonretardedBulk}(a), the 2DEG lies in the $z=0$ plane. The $0<z<\DdA$ region is filled with an insulator (or vacuum) of the dielectric constant $\EpsA$. The $-\DdB<z<0$ region is filled with another insulator (or semiconductor) of the dielectric constant $\EpsB$. The $z>\DdA$ and $z<-\DdB$ regions are filled with perfect metals whose dielectric constant is $\EpsM=-\infty$ in the (radio to microwave) frequency range of MP. Encapsulating the system by two metals mimics the experimental configurations with top and bottom electrodes, and in the meanwhile, cutoffs the theoretically infinitely long-ranged Coulomb interaction.

For the results shown in the main article, we simply choose $\EpsA=\EpsB=\epsilon_0=1$ (vacuum permittivity), $m_*=m_0$ (bare electron mass), $\DdA=\DdB=d=1~\mu$m, and $n_0=1~\mu$m$^{-2}=10^{8}$~cm$^{-2}$. We can thus obtain a characteristic frequency, $\omega_\star=\sqrt{\frac{2\pi e^2n_0}{m_0d}}=3.99\times10^{10}$~s$^{-1}$, (i.e., $f_\star=\frac{\omega_\star}{2\pi}=6.35$~GHz), which is used to normalize the frequencies. For the case of cyclotron frequency $\OmegaC=\omega_\star$, we have $B_0=0.227$~T.

\ \\\noindent\textsf{\textbf{Cartesian representation.}}
In the main article we have used the chiral representation for the bulk Hamiltonian and generalized current vector, in order to demonstrate the similarity between our 2D MP and the 2D $p+\Ii p$ topological superconductor. But it is sometimes more convenient to adopt the traditional Cartesian representation, especially when we solve for the edge modes and apply boundary conditions on the edge. In the Cartesian representation, the bulk Hamiltonian equation reads
\[
\omega \mathbf{J}(\mathbf{q},\omega) = \mathcal{H}(\mathbf{q}) \mathbf{J}(\mathbf{q},\omega) ,\quad \mathbf{J}(\mathbf{q},\omega)\equiv
\begin{pmatrix}
\calJ_x(\mathbf{q},\omega)\\
\JD(\mathbf{q},\omega)\\
\calJ_y(\mathbf{q},\omega)
\end{pmatrix},
\]
where the Hamiltonian in this representation reads
\begin{align}
\mathcal{H}(\mathbf{q})
&=
\begin{pmatrix}
0 & \DS\frac{\OmegaP(q)}{q} q_x & -\Ii\OmegaC \\
\DS\frac{\OmegaP(q)}{q} q_x & 0 & \DS\frac{\OmegaP(q)}{q} q_y \\
+\Ii\OmegaC & \DS\frac{\OmegaP(q)}{q} q_y & 0
\end{pmatrix} \nonumber \\
&\overset{q\rightarrow0}{\longrightarrow}
\begin{pmatrix}
0 & \vP q_x & -\Ii\OmegaC \\
\vP q_x & 0 & \vP q_y \\
+\Ii\OmegaC & \vP q_y & 0
\end{pmatrix}.
\label{EqnHamiltonianOld}
\end{align}
The particle-hole and time-reversal operators in this Cartesian representation are
\begin{equation}
\mathcal{C}=
\left.\begin{pmatrix}
1&0&0\\
0&1&0 \\
0&0&1
\end{pmatrix}
K \right|_{\mathbf{q}\rightarrow-\mathbf{q}},
\quad
\mathcal{T}=
\left.\begin{pmatrix}
-1&0&0\\
0&1&0 \\
0&0&-1
\end{pmatrix}
K \right|_{\mathbf{q}\rightarrow-\mathbf{q}}.
\end{equation}
Clearly, $\mathcal{C}$ is just a complex conjugation on the matrix or vector elements (combined with a momentum reversal $\mathbf{q}\rightarrow-\mathbf{q}$ here, when written on the single-particle momentum bases).

\ \\\noindent\textsf{\textbf{Numerical scheme.}} To numerically solve the nonuniform edge-state problem, which is unrestricted by the long-wavelength approximation, we expand the governing Eqs.~(\ref{EqnContinuity}--\ref{EqnEOM}) in the main article into plane waves and use the form of interaction in the momentum space Eq.~(\ref{EqnNonretardedField}). We can derive a matrix equation in the Cartesian representation,
\begin{equation}
\begin{split}
\omega
\begin{pmatrix}
\mathbf{J}_x(\mathbf{q},\omega) \\
\mathbf{J}_{\SSS\text{D}}(\mathbf{q},\omega) \\
\mathbf{J}_y(\mathbf{q},\omega)
\end{pmatrix}
=&
\begin{pmatrix}
0 & \DS \mathbf{U}_x(\mathbf{q};\mathbf{q}') & \DS -\Ii \mathbf{W}(\mathbf{q};\mathbf{q}') \\
\DS \mathbf{Q}_x(\mathbf{q};\mathbf{q}') & 0 & \DS \mathbf{Q}_y(\mathbf{q};\mathbf{q}') \\
\DS +\Ii \mathbf{W}(\mathbf{q};\mathbf{q}') & \DS \mathbf{U}_y(\mathbf{q};\mathbf{q}') & 0
\end{pmatrix} \\
&
\begin{pmatrix}
\mathbf{J}_x(\mathbf{q}',\omega) \\
\mathbf{J}_{\SSS\text{D}}(\mathbf{q}',\omega) \\
\mathbf{J}_y(\mathbf{q}',\omega)
\end{pmatrix}
\end{split}. \label{EqnComplete}
\end{equation}
The sub-matrices $\mathbf{U}_x(\mathbf{q};\mathbf{q}')$, $\mathbf{U}_y(\mathbf{q};\mathbf{q}')$, $\mathbf{Q}_x(\mathbf{q};\mathbf{q}')$, $\mathbf{Q}_y(\mathbf{q};\mathbf{q}')$, $\mathbf{W}(\mathbf{q};\mathbf{q}')$, and the sub-column-vectors $\mathbf{J}_x(\mathbf{q}')$, $\mathbf{J}_{\SSS\text{D}}(\mathbf{q}')$, $\mathbf{J}_y(\mathbf{q}')$ all have the dimension of the number of plane waves used for expansion.

The elements of sub-matrices are
\begin{align}
\mathbf{U}_x(\mathbf{q};\mathbf{q}') &= 2\pi\tilde{\alpha}(\mathbf{q}-\mathbf{q}') \frac{q_x'}{q'\xi(q')}, \\
\mathbf{U}_y(\mathbf{q};\mathbf{q}') &= 2\pi\tilde{\alpha}(\mathbf{q}-\mathbf{q}') \frac{q_y'}{q'\xi(q')},
\end{align}
\begin{align}
\mathbf{Q}_x(\mathbf{q};\mathbf{q}') &= \delta_{\mathbf{q},\mathbf{q}'} q_x', \\
\mathbf{Q}_y(\mathbf{q};\mathbf{q}') &= \delta_{\mathbf{q},\mathbf{q}'} q_y',
\end{align}
and
\begin{align}
\mathbf{W}(\mathbf{q};\mathbf{q}') &= \OmegaC(\mathbf{q}-\mathbf{q}').
\end{align}
where $\tilde{\alpha}(\Delta \mathbf{q})$ and $\OmegaC(\Delta\mathbf{q})$ with $\Delta\mathbf{q}=\mathbf{q}-\mathbf{q}'$ are the Fourier transform of $\alpha(\mathbf{r})$ and $\OmegaC(\mathbf{r})$. For the system with edges along the $y$-axis, the expansion only needs to be done in the $x$-direction while keeping $q_y=q_y'$ as good quantum numbers. Typically, we use 128 plane waves to do the expansion, and the standard eigen-solver to obtain all the eigenfrequencies and eigenvectors.

The 2D real-space calculation for Fig.~\ref{FigMovie} is performed by a homemade finite-difference time-domain code. The field quantities are defined on a square grid. Eqs.~(\ref{EqnContinuity}--\ref{EqnEOM}) are adopted for the real-time evolution after replacing $-\Ii\omega$ with $\partial_t$. A point-source term of a given driving frequency is added to the right-hand side of Eq.~(\ref{EqnEOM}).

\ \\\noindent\textsf{\textbf{Data availability.}} All relevant data are available from the authors on request.

}

\ \\\noindent\textsf{\textbf{Acknowledgements}}

\noindent We would like to thank Duncan Haldane and Aris Alexandradinata for discussing the reality condition. We are also thankful to Patrick A. Lee, Xiaogang Wen, Frank Wilczek, Yang Qi, Junwei Liu, and Fan Wang for helpful discussions. D. J. and N. X. F. were supported by AFOSR MURI under Award No. FA9550-12-1-0488. L. L. was supported in part by the MRSEC Program of the NSF under Award No. DMR-1419807, the National Thousand-Young-Talents Program of China, the National 973 Program of China (Nos. 2013CB632704 and 2013CB922404), and the National Natural Science Foundation of China (No. 11434017). L. L. and C. F. were supported in part by the Ministry of Science and Technology of China under Grant No. 2016YFA0302400. Z. W. was supported by NSFC under Grant No. 11674189. C. F. and L. F. were supported by the DOE Office of Basic Energy Sciences, Division of Materials Sciences and Engineering under Award No. DE-SC0010526. J. D. J. were supported in part by the US ARO through the ISN, under Contract No. W911NF-13-D-0001. M. S. and L. L. were supported in part by the MIT S3TEC EFRC of DOE under Grant No. DE-SC0001299.

\ \\\noindent\textsf{\textbf{Author contributions}}

\noindent D. J. and L. L. conceived the concept and wrote the manuscript. D. J. performed the analytical and numerical calculation. Z. W.,  C. F. and L. F. joined effective discussion and provided pivotal advices. J. D. J. and M. S. offered additional guidance and proofread the manuscript. L. L. and N. X. F. directed the project.

\bibliography{References}

\end{document}